\documentclass[runningheads]{llncs}
\usepackage{graphicx}

\usepackage{cite}      
\usepackage[utf8]{inputenc} 
\usepackage[T1]{fontenc}    
\usepackage{hyperref}       
\usepackage{url}            
\usepackage{booktabs}       
\usepackage{amsfonts}       
\usepackage{nicefrac}       
\usepackage{microtype}      
\usepackage{xcolor}         
\usepackage{amsmath}
\usepackage{amssymb}
\usepackage{lipsum}
\usepackage{subfigure}
\usepackage{graphicx}
\usepackage{threeparttable}
\usepackage{multirow}

\begin{document}
\title{Macro2Micro: A Rapid and Precise Cross-modal Magnetic Resonance Imaging Synthesis using Multi-scale Structural Brain Similarity}

\author{Sooyoung Kim\inst{1,}$^*$ \and Joonwoo Kwon\inst{2,}$^*$\and Junbeom Kwon\inst{3,}$^*$ \and Jungyoun Janice Min\inst{1} \and Sangyoon Bae \inst{4} \and Yuewei Lin\inst{5} \and Shinjae Yoo\inst{5}\and Jiook Cha\inst{3,4,}$^\dagger$}

\authorrunning{S.Kim et al.}

\institute{Department of Brain and Cognitive Science, Seoul National University, Seoul, Republic of Korea \and
Department of Applied Bioengineering, Seoul National University, Seoul, Republic of Korea \and
Department of Psychology, Seoul National University, Seoul, Republic of Korea \and
Interdisciplinary Program in Artificial Intelligence, Seoul National University, Seoul, Republic of Korea \and
Computational Science Initiative, Brookhaven National Laboratory, Upton, NY, USA}

\maketitle     

\def\thefootnote{*}\footnotetext{These authors contributed equally to this work.}
\def\thefootnote{$\dagger$}\footnotetext{Corresponding Author.}

\begin{abstract}
The human brain is a complex system requiring both macroscopic and microscopic components for comprehensive understanding. However, mapping nonlinear relationships between these scales remains challenging due to technical limitations and the high cost of multimodal Magnetic Resonance Imaging (MRI) acquisition. To address this, we introduce \textbf{Macro2Micro}, a deep learning framework that predicts brain microstructure from macrostructure using a Generative Adversarial Network (GAN). Based on the hypothesis that microscale structural information can be inferred from macroscale structures, Macro2Micro explicitly encodes multiscale brain information into distinct processing branches. To enhance artifact elimination and output quality, we propose a simple yet effective auxiliary discriminator and learning objective. Extensive experiments demonstrated that Macro2Micro faithfully translates T1-weighted MRIs into corresponding Fractional Anisotropy (FA) images, achieving a 6.8\% improvement in the Structural Similarity Index Measure (SSIM) compared to previous methods, while retaining the individual biological characteristics of the brain. With an inference time of less than 0.01 seconds per MR modality translation, Macro2Micro introduces the potential for real-time multimodal rendering in medical and research applications. The code will be made available upon acceptance.

\end{abstract}

\section{Introduction}
\label{introduction}

At a large scale, the human brain is composed of multiple components. The macrostructure of the brain encompasses the architecture of cell bodies and neurites in the gray matter, as well as the arrangement and myelin content of axons in the white matter. It also includes the presence of cerebrospinal fluid (CSF), fat, and inflammation. Understanding the macro-structural aspects of the brain is essential for comprehending its structure and organization. This knowledge provides valuable information for detecting abnormalities associated with disorders such as lesions\cite{ginat2012intracranial} and neurodegenerative disease\cite{ harper2016mri, frisoni2010clinical}, and multiple sclerosis \cite{rocca2017role}. On the other hand, the brain's microstructure consists of minuscule components such as myelin, axons, and dendrites, with a particular emphasis on their characteristics such as cell size, fiber orientation, and density\cite{ranzenberger2019diffusion}. It plays a crucial role in identifying tissue damage and anomalies associated with cell density\cite{kinoshita2008fractional}, and in studying psychiatric illnesses like schizophrenia\cite{kubicki2007review} and depression\cite{qin2015altered}, as well as neurological diseases such as Alzheimer's dementia\cite{douaud2011dti, mayo2017longitudinal} and Parkinson's Disease\cite{zheng2014dti, ofori2015longitudinal}.




With the advent of Magnetic Resonance Imaging (MRI), the macrostructure of the human brain could be captured by structural MRIs (sMRI), such as T1-weighted MRI. These images primarily depict the shape and the size of brain tissue\cite{chen2018t1}. In contrast, models that are based on diffusion MRI (dMRI), namely Diffusion-Tensor Imaging (DTI), are mainly employed for the examination of microstructures\cite{gu2019generating}. DTI is used to parametrize diffusion of water molecules, aiding in estimating properties of tissue microstructure and identifying the brain's white fiber tracts.

Studying both the macro- and micro-structures of the brain is crucial for a comprehensive understanding of its complex connectivity and function, as they provide complementary information \cite{dyrba2012combining, li2013discriminative}. There have been attempts to integrate modalities encoding different structural scales of human brain information in a multimodal fashion \cite{vasung2013multimodality, liang2019classificatio}. Nonetheless, obtaining such MRI modalities is \textit{extremely} costly and time-consuming. For instance, T1-weighted MRI requires a single scan, while DTI requires multiple scans, at least six for accurate measurements and ideally thirty scans for rotational invariance \cite{jones2004effect,jones2013white}. In addition, obtaining different MRI modalities \textit{simultaneously} in high quality requires a lengthy acquisition time. It can cause patients with panic disorder or agoraphobia due to the potentially stressful nature of the MRI environment\cite{lueken2011don, murphy1997adult}, making it practically impossible to obtain them at the same time in a clinical manner. Moreover, the long duration of MRI scans can cause motion artifacts \cite{rzedzian1983real, tsao2010ultrafast}, a problem that has been continuously raised in the context of time-consuming DTI \cite{aksoy2008single}. Despite these challenges, DTI's clinical significance remains, and efforts are ongoing to obtain high-quality DTI with fewer scans \cite{tian2020deepdti}, though the need for multiple scans persists.

In the meantime, T1-weighted sMRIs, which primarily capture the macrostructure of the brain, offer insights into the precise positioning and dispersion of white fiver bundles throughout the brain and even provide microstructural data comparable to that obtained through DTI. T1-weighted sMRIs can estimate the location, distribution, and thickness of white fiber tracts, essential elements in constructing DTI, by analyzing the correlation between signal intensity and the density of these bundles. T1-weighted sMRI is a great resource for generating DTI and comprehending brain microstructure, as evidenced by studies involving cross-species\cite{harkins2016microstructural}. This demonstrates the potential of macrostructure in facilitating the creation of microstructure in the brain. By taking these insights into account during synthesis, one can produce more favorable outcomes.

Moreover, in a deep learning approach where a large amount of data is crucial, only subjects with data for all modalities can be used for the analysis. As a result, the available data size is inevitably reduced, leading to poor generalizability and an increased risk of overfitting. Hence, it is imperative to develop fast, accurate, and robust cross-modal MRI-generating techniques to fully understand the brain comprehensively and overcome the limited data availability.

Nevertheless, the existing methods\cite{Yang2020CMI2I, sikka2021mri} struggle to effectively translate target MRI modalities with satisfactory quality, and their capabilities are restricted to translating within sMRIs. Moreover, they often neglect the importance of preserving the individual biological characteristics inherent to the images. The academic and clinical applicabilities of the generated image are greatly restricted if it just imitates the distribution of training data, notwithstanding its potential utility.

This study proposes a new image-to-image translation framework for a cross-MRI modality synthesis. We named our framework \textbf{Macro2Micro} as it operates under the assumption that the macroscale structure of the brain can provide insights into its microscale structure. Macro2Micro utilizes Generative Adversarial Network (GAN)\cite{goodfellow2020generative} to encode multiscale structural information into separate processing branches. During the encoding and decoding process, an active information exchange occurs between two frequency processing branches. This interchange allows the model to capture structural connectivity between two different MRI modalities. 
We thoroughly compare and validate the superiority of our proposed model against conventional image-to-image translation studies and achieve the highest quality, both qualitatively and quantitatively. We also explore the possibility of generating tractography from DTI using our method. This enables us to explore the potential of our model in capturing the structural connectivity information of the brain and generating diverse target MRI modalities while retaining individual biological characteristics.

\section{Method}
\label{sec:networks}

\begin{figure}[t]
    \centering
    \includegraphics[width=\textwidth]{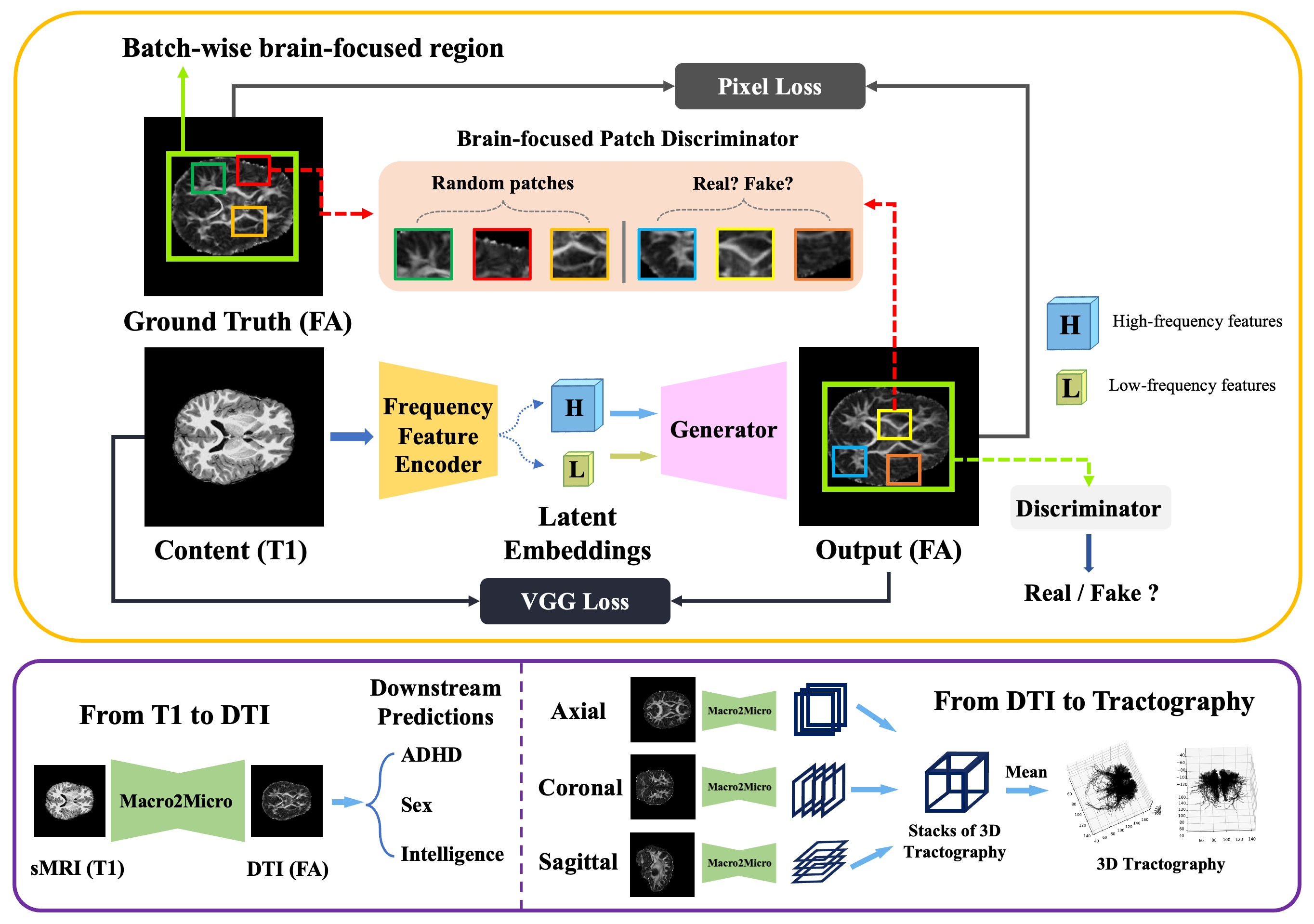}
    \caption{The overall architecture of the proposed model. (Top) and its versatility we experimented in this paper. (Bottom)}
    \label{fig:model_architecture}
\end{figure}

\begin{figure}[t]
    \centering
    \includegraphics[width=\textwidth]{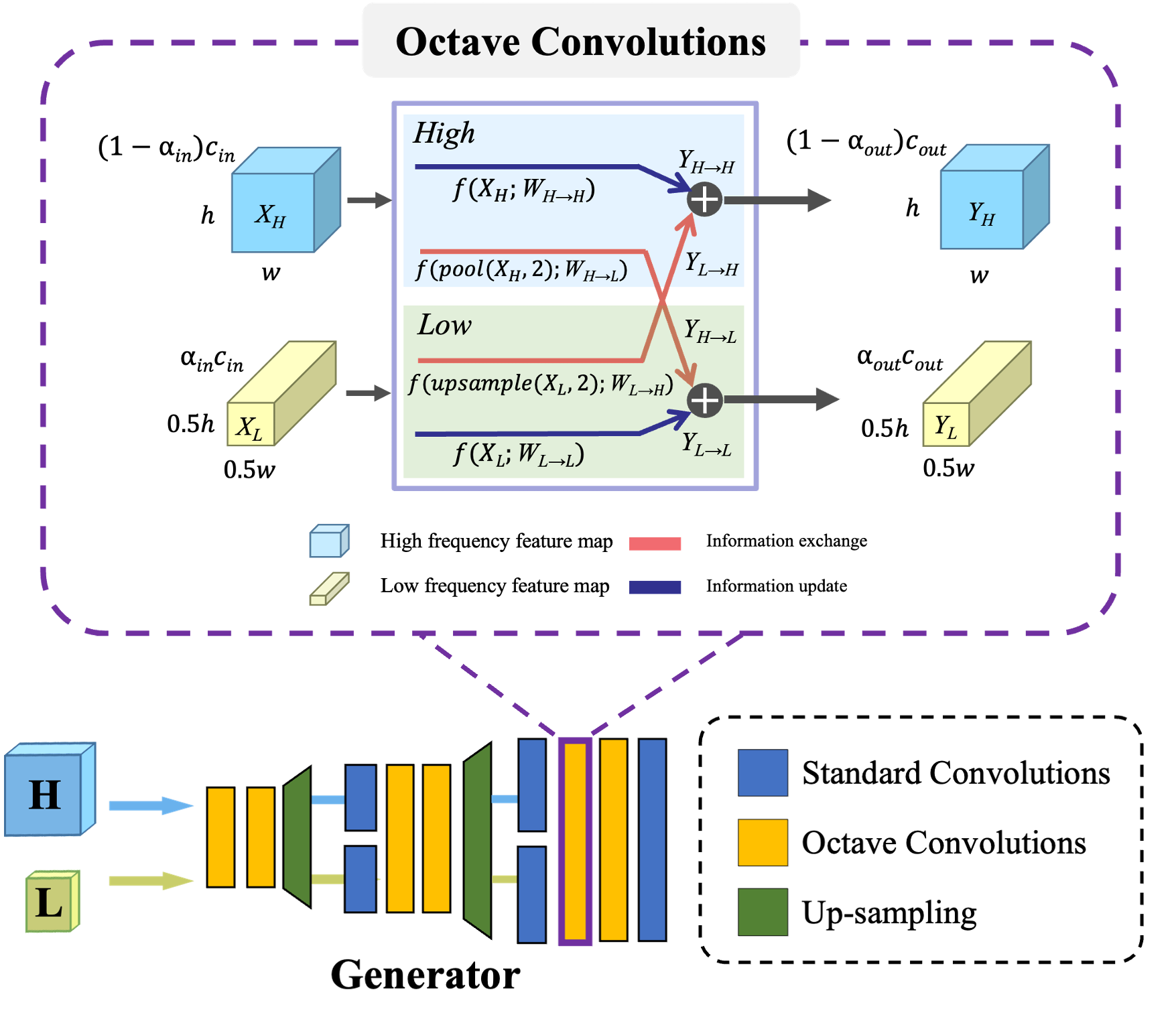}
    \caption{The structure of Octave Convolutions (Top) and the generator design (Bottom). \textit{H} and \textit{L} represent High- and Low-frequency features respectively.
    }
    \label{fig:octave_architecture}
\end{figure}

Fig.~\ref{fig:model_architecture} gives an overview of our framework for cross-modal MRI synthesis (yellow outline) alongside its versatile utilities outlined in our paper (purple outline). The subsequent sections include an exhaustive analysis of the proposed method and its underlying principles. 

\textbf{Architecture Overview.} The proposed architecture consists of four components: a frequency feature encoder $E$, a generator $G$, a discriminator $D$, and a brain-focused patch discriminator $D_{brainPD}$. Specifically, the input MR images are decomposed into two distinct frequency feature maps through a frequency feature encoder $E$, encoding information on the macro-structure of the brain into low-frequency components and information on the microstructure of the brain into the high-frequency components. These encoded latent features are subsequently fed to the generator, where the synthesized high and low-frequency outputs of the target modality are made from the input latent features. The terminal layer of the generator combines the synthesized high and low-frequency outputs to generate the final output. This output are then fed to both the discriminator $D$ and the brain-focused patch discriminator $D_{brainPD}$. These two discriminators guide our model to effectively synthesize target modality by focusing on the delicate details in the brain region in the input and learning the statistical relationships between each image patch.

\textbf{Octave Convolutions.} One of the key elements of the proposed model is utilizing Octave Convolutions (OctConv)\cite{chen2019drop} to encode macro- and micro-scale information into the corresponding frequency feature maps. To factorize input mixed feature maps based on their frequencies, the spatial resolution of low-frequency feature maps in OctConv is decreased by one octave, where the term \textit{octave} refers to a spatial dimension divided by a power of two. In this study, a value of 2 was chosen for simplicity. The spatial reduction in the low-frequency branch expands the receptive field of the low-frequency processing branch, capturing more contextual information from distant locations and improving synthesis performance. Following the convention presented by previous research \cite{kwon2024aesfa}, the up-sampling order is modified to effectively address checkerboard artifacts \cite{odena2016deconvolution}. The detailed design of OctConv used in this study is illustrated in Fig.~\ref{fig:octave_architecture}. Here, the $\alpha$ value refers to the ratio of the low-frequency channels to high-frequency channels, and empirical findings indicate that employing OctConv with half the channels for each frequency ($\alpha=0.5$) yields optimal performance. A comprehensive experiment regarding this matter can be found in the section~\ref{sec:result_octave}.

\textbf{Frequency Feature Networks.} Both the frequency feature encoder and the generator are equipped with several layers of OctConvs. This idea was originated from the previous study \cite{kwon2024aesfa}. However, our work differs from theirs in that we primarily focus on encoding different scales of information from the brain into the corresponding frequency components without any auxiliary encoder. While latent features decomposed by the encoder are convolved in the generator, the two frequency features actively exchange information with the opponent via information exchange branches. This active information exchange between frequency components compensates for the missing information in each branch and boosts the entire synthesis process. The standard convolutions after up-sampling operations are responsible for learning frequency-agnostic information and compensating for the missing information during up-sampling operations. The effectiveness of active information exchange is outlined in the section~\ref{sec:result_octave}. 

\textbf{Brain-focused Patch Discriminator.} While using the discriminator solely seems sufficient for synthesizing the target modality, the results still suffer from the checkerboard artifacts and undesired artifacts (see the section~\ref{sec:ablations} for the details). To tackle this, we employed a patch co-occurrence discriminator, introduced by \cite{park2020swapping}. We encourage patches cropped from the output to maintain the identical representation as the patches cropped from the target MR images. Consequently, the generator aims to generate an output image such that any patch from the output cannot be distinguished from a group of patches from the actual MR images. 

However, most Magnetic Resonance brain Images contain redundant background regions. These regions are primarily zero-values or noises. Cropping patches from such regions and feeding them to the discriminator are inefficient and could lead to the degradation of output image quality (e.g., blurring or dimmer images and pixelization) as the model would learn the background noises or the abrupt changes in the boundary of our brain and the background. To effectively cope with this, we applied a simple yet effective pre-processing algorithm. We first calculate the valid brain regions in the training mini-batch, which are then used to crop the valid region from the given training mini-batch. By doing so, our brain-focused patch discriminator serves to focus on the effective regions of the brain and enforce that the joint statistics of a learned representation consistently follow the ground truth modality. 

\textbf{Learning Objectives.} To guide our model to learn subject-independent representation and the connectivity between the macro and micro-structure of the human brain while synthesizing the target modality with desired image quality, we use the mean square error ($\mathcal{L}_{\text{pix}}$) between the output $I_{out}$ and ground truth $I_{GT}$ and the discriminator objectives ($\mathcal{L}_{\text{GAN}}$):

\begin{equation*}
\begin{gathered}
\mathcal{L}_{\text{pix}}=\|I_{\text{out}}-I_{\text{GT}}\|_{\text{1}} , \\
\mathcal{L}_{\text{GAN}}=\mathbb{E}[-log(D(I_{\text{out}}))]
\label{eq:pix_loss}
\end{gathered}
\end{equation*}

For the brain-focused patch discriminator, we follow the loss of Swap-AE \cite{park2020swapping}, but with slight changes described in previous section. The final GAN loss for the brain-focused patch discriminator is as follows:
\begin{equation*}
\mathcal{L}_{\text{patch}}=\mathbb{E}[-log(D_{\text{patch}}(crops(valid(I_{\text{out}})), crops(valid(I_{\text{GT}}))))]
\label{eq:patch_loss}
\end{equation*}
where \textit{crops} operator selects a random patch of size $1/2$ to $1/3$ of the full image dimension on each side and \textit{valid} operator calculates the valid brain regions in the given training mini-batch and then crops according to them.

To prevent the model from falling the mode-collapse and generating skull-like artifacts (see the details in Fig.~\ref{fig:brainPD_percept_loss_distance}), we utilize prior knowledge from a pre-trained convolutional neural network, such as VGG-19 \cite{simonyan2014very}. The perceptual loss was originally proposed by \cite{johnson2016perceptual}, yet has not been actively addressed in the Magnetic Resonance Imaging domain to cope with the mode collapse. The perceptual objective we used is as follows:

\begin{equation*}
\mathcal{L}_{\text{perct}}=\sum_{n=1}^{4}\|f_{\text{n}}(I_{\text{out}})-f_{\text{n}}(I_{\text{GT}})\|_{\text{2}}
\label{eq:perct_loss}
\end{equation*}
where $f_{n}$ symbolizes the \textit{n}-th layer in the VGG-19 model. The perceptual loss is computed at the \{\textit{conv1\_1, conv2\_1, conv3\_1, conv}4\_1\}.
Considering all the aforementioned losses, the total loss is formalized as:
\begin{equation*}
\mathcal{L}_{\text{total}}=\lambda_{\text{pix}}\mathcal{L}_{\text{pix}}+\lambda_{\text{perct}}\mathcal{L}_{\text{perct}}+\lambda_{\text{GAN}}\mathcal{L}_{\text{GAN}}+\lambda_{\text{patch}}\mathcal{L}_{\text{patch}}
\label{eq:train_loss}
\end{equation*}
where $\lambda_{\text{pix}}$, $\lambda_{\text{perct}}$, $\lambda_{\text{GAN}}$, and $\lambda_{\text{patch}}$ are the weighting hyper-parameters for each loss.

\textbf{Experimental Settings and Data.} To show the versatility of the proposed model, we designed two separate experimental settings. First, we synthesize the Diffusion Tensor Image (DTI) from the sMRI (Left side of purple outline in Fig.\ref{fig:model_architecture}). We used Fractional Anisotropy (FA) for DTI and T1-weighted images for sMRI. Next, we synthesize the tractography from the FA images (Right side of purple outline in Fig.\ref{fig:model_architecture}). To synthesize 3D tractography, we trained three different 2D synthesis models that translate 2D tractography using FA images sliced from three distinct view (axial, coronal, and sagittal view of the brain). Then, we collected all three stacks of 2D tractography from each trained model and merged using the \textit{Mean} operation to generate the final 3D tractography. 

We use the Adolescent Brain Cognitive Development (ABCD) dataset \cite{casey2018adolescent}, which consists of comprehensive developmental data including structural brain MRI in children across various sites in the United States. The data utilized in this study is obtained from ABCD, and the image acquisition protocol and minimal processing pipeline can be found in previous studies \cite{casey2018adolescent, hagler2019image}. For image-to-image translation between T1-weighted images and DTI, the T1 image and fractional anisotropy (FA) for DTI have dimensions of $256 \times 256 \times 256$ and a voxel size of 1mm. We use quality controlled 7,669 subjects. For translation between FA and tractography, the FA image has dimensions of 190 $\times 190 \times 190$ and a voxel size of 1.25mm. The analysis includes a total of 6,365 subjects. Probabilistic tractography was performed using 1 million tracks derived from DTI data. Only the streamlines passing by vmPFC (ventromedial prefrontal cortex) based on the aparc+aseg atlas\cite{fischl2002whole} is extracted, and the shape is adjusted to match the FA image.

\textbf{Implementation Details.}
During training, all images are loaded as 256x256 pixels and scaled to [0, 1]. The model is trained using the Adam optimizer \cite{kingma2014adam} with a learning rate 0.0002 and a batch size of 8 for 200 epochs. The encoder feature map has dimensions of (128, 64, 64) for high and (128, 32, 32) for low-frequency components. The baseline models outlined in this paper were trained using the author-released codes and parameters. Baselines and all our experiments are conducted using the PyTorch framework \cite{paszke2019pytorch} on a single NVIDIA RTX A5000(24G) GPU.

\section{Results}

\begin{figure*}[t!]
    \centering
    \includegraphics[width=\textwidth, height=6cm]{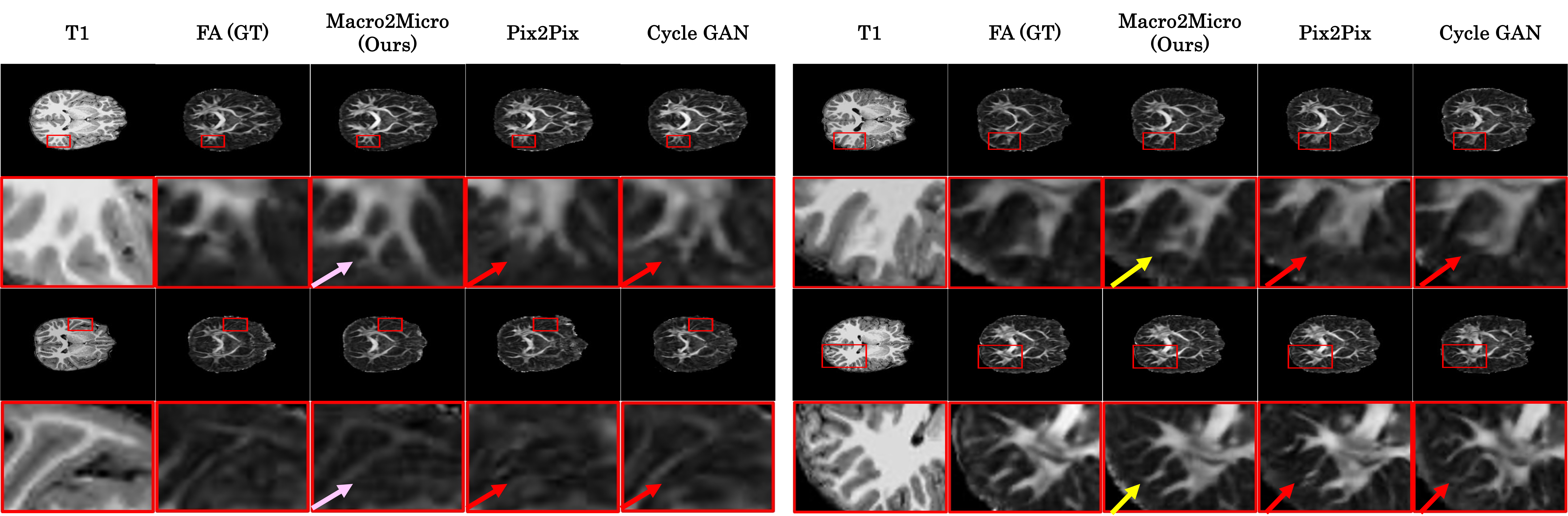}
    \caption{Qualitative Comparison of images made with our proposed model (Macro2Micro), and images made with Pix2Pix and CycleGAN. Magnify the image to see the details.}
    \label{fig:Baseline}
\end{figure*}

\begin{table}[t!]
\small \centering
\caption{Quantitative comparison with generated whole brain FA (FA) images and white matter (WM) whose FA value is bigger than 0.2 from synthesis outputs. The best outcomes are shown in \textbf{bold}. ↑: Higher is better. ↓: Lower is better.}
\label{tab:quantitative_comparison}
\begin{tabular}{cccccc}
\hline
Methods                     & Input (GT)     &SSIM ($\uparrow$) &PSNR ($\uparrow$) &MAE ($\downarrow$)&MSE($\downarrow$)            \\ \hline
Pix2Pix                     &  FA         & 0.8310          & 24.7738          & 0.1469          & 0.1292          \\
CycleGAN                    &  FA         & 0.8332          & 24.6660          & 0.1477          & 0.1299          \\ 
Macro2Micro (Ours) & FA & \textbf{0.8600} & \textbf{25.7560} & \textbf{0.1383} & \textbf{0.1226} \\ \hline \hline

Pix2Pix                     &  MD         &           &           &           &           \\
CycleGAN                    &  MD         &           &           &           &           \\ 
Macro2Micro (Ours) & MD & \textbf{} & \textbf{} & \textbf{} & \textbf{} \\ \hline \hline

Pix2Pix                     &  WM         & 0.8354          & 24.8992          & 0.1374          & 0.1200          \\
CycleGAN                    &  WM         & 0.8369          & 24.7585          & 0.1383          & 0.1208          \\ 
Macro2Micro (Ours) & WM & \textbf{0.8627} & \textbf{25.8493} & \textbf{0.1300} & \textbf{0.1146} \\ \hline
\end{tabular}
\end{table}

\textbf{Diffusion Tensor Image Synthesis.} We compared our Macro2Mciro model with existing image translation models (i.e., Pix2Pix \cite{isola2017image} and CycleGAN \cite{zhu2017unpaired}) (Fig.~\ref{fig:Baseline}). Our model could faithfully reconstruct the structural location and FA value with minimum residuals while maintaining both isotropic and anisotropic movement of water molecules in the given brain. In terms of white matter and overall brain structure, all models translate target modality with comparable quality. However, Pix2Pix and CycleGAN could not recognize and synthesize intricate micro-scale structures in many cases. For instance, results from baseline models mostly neglected and underexpressed the microscopic white matter at the boundaries of the brain (red arrows), whereas our model does not (green arrows). Owing to its ability to learn both macro- and microstructures from both modalities, our model not only generates the most comparable images to the ground truths but also captures structural details lost in the original and reconstructs parts that were previously disconnected or absent. For example, our model learns the presence of the white matter from FA image while determining its morphology from the T1 image (pink and yellow arrows and their corresponding T1 and FA). Similarly, our model shows better quality by bending the straight white matter line in ground truth to its more specific endpoint (yellow arrows) by referencing the macro-structure from the T1 image. In terms of quantitative comparisons, Macro2Micro shows the best performance, achieving the best SSIM, PSNR, MAE, and MSE (Table~\ref{tab:quantitative_comparison}). In addition, this result holds even when we compared the results within the white matter regions (voxels with FA greater than 0.2).

\begin{figure}[t]
    \centering
    \includegraphics[width=\textwidth]{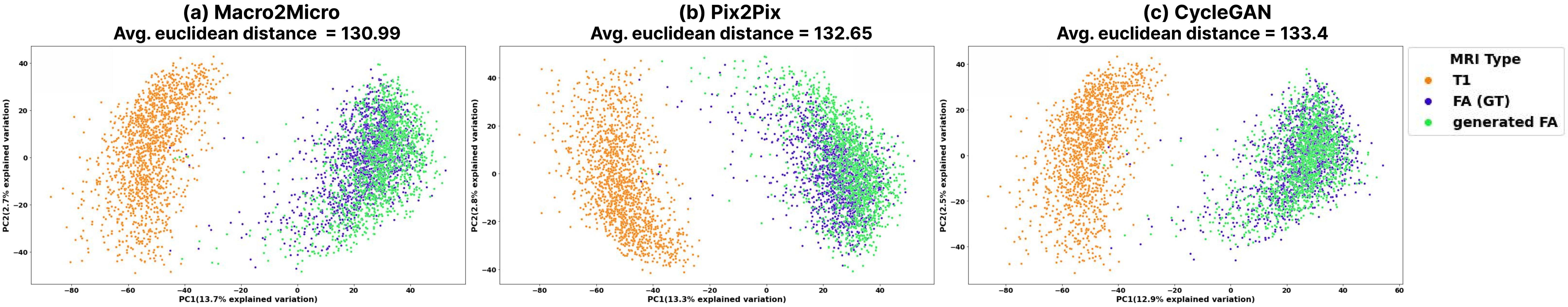}
    \caption{PCA results for the T1, FA (\textbf{G}round \textbf{T}ruth), and generated FA images from (a) Micro2Macro, (b) Pix2Pix, and (c) CycleGAN. The Euclidian distance below each model name indicates the average distance between the FA (Ground Truth) and the generated FA images.}
    \label{fig:PCA_macro2micro}
\end{figure}

To further test the robustness and effectiveness of the proposed model, we conducted principal component analysis (PCA) on three types of images: T1-weighted images, ground truth FA images, and synthesized FA images. We analyzed and visualized 1,499 subjects in the test dataset. After flattening $256^2$ voxels from MRI slices, we removed the non-brain background, leaving 26,891 voxels for each brain modality, PCA was then applied to the entire features. A batch size of 200 was utilized for the incremental PCA and the visualization was performed using two principal components. The two principal components individually accounted for 13.7\% and 2.7\% of the variances. Notably, the PCA result demonstrated that the generated FA images exhibited significant dissimilarity compared to the original T1 images while displaying an overlap with the ground truth FA images in the low-dimensional representation (Fig.~\ref{fig:PCA_macro2micro}).

\textbf{Prediction of Biological and Cognitive Variables.} The image-to-image translation approach is powerful in terms of image transformation; however, there may be a major concern about potential damage to biological characteristics~\cite{cohen2018distribution}. To test whether biological information from the brain images is preserved during the translation process, we conducted a task of predicting the sex, intelligence, and Attention-Deficit/Hyperactivity Disorder (ADHD) diagnosis of children using both predicted and ground truth FA images. Implementation details for these downstream tasks are in supplementary materials. We present the performance of the generalized linear model (GLM) on sex, intelligence, and ADHD prediction tasks in Table~\ref{tab:downstream_t1_fa}. It is worth noting that Macro2Micro yielded an AUROC of 0.782 in the sex classification, slightly higher than that of ground-truth FA images (0.7641) and other algorithms. Although Pix2Pix showed the best performance in predicting ADHD diagnosis and intelligence, our model exhibited comparable performance for predicting ADHD diagnosis to ground truth FA images in terms of AUROC and surpassed the ground truth FA in predicting Intelligence. This suggests that the biological characteristics of different individuals could be preserved.

\begin{table}[t]
\small \centering
\caption{ADHD, Sex classification and Intelligence regression performance of real T1, real FA, and synthesized FA images.}
\label{tab:downstream_t1_fa}
\begin{tabular}{clcccccc}
\multicolumn{2}{c}{\multirow{2}{*}{Input}} & \multicolumn{2}{c}{ADHD}          & \multicolumn{2}{c}{Sex}             & \multicolumn{2}{c}{Intelligence} \\ \cline{3-8} 
\multicolumn{2}{c}{}                       & AUROC ($\uparrow$)   & ACC ($\uparrow$)   & AUROC ($\uparrow$)  & ACC ($\uparrow$)  & Corr.Coef.  ($\uparrow$)     & MSE ($\downarrow$)     \\ \hline
\multicolumn{2}{c}{T1}                     & 0.5034          & 0.5479          & 0.7820           & 0.7133           & 0.159           & 0.777          \\
\multicolumn{2}{c}{FA (GT)}                & 0.4812          & 0.5342          & 0.7641           & 0.7066           & 0.124           & 0.832          \\
\multicolumn{2}{c}{FA (Pix2Pix)}           & \textbf{0.5532} & \textbf{0.5821} & 0.7565           & 0.6800           & \textbf{0.187}  & \textbf{0.784} \\
\multicolumn{2}{c}{FA (CycleGAN)}          & 0.4445          & 0.4794          & 0.7534           & 0.6766           & 0.066           & 0.836          \\ \hline
\multicolumn{2}{c}{FA (Macro2Micro)}       & 0.4926          & 0.5136          & \textbf{0.7726} & \textbf{0.6866} & 0.166  & 0.797
\end{tabular}
\end{table}

\begin{figure}[t]
    \centering
    \includegraphics[height=7cm]{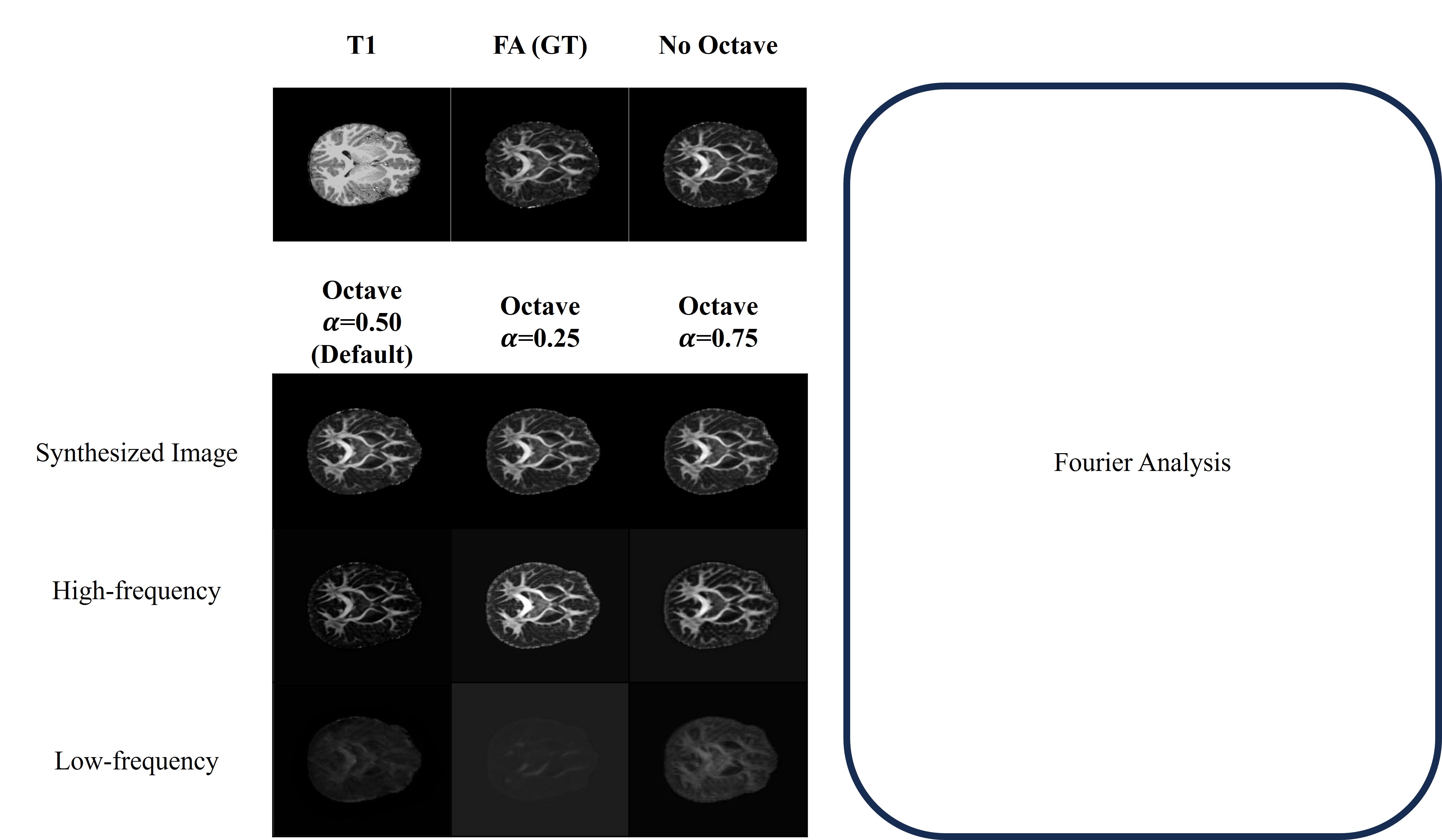}
    \caption{\textbf{Ablation study} Qualitative comparison of the effectiveness of Octave Convolution.}
    \label{fig:Octave}
\end{figure}

\begin{table}[t!]
\small \centering
\caption{Effects of brain-focused patch discriminator, the utilization of perceptual loss with a pre-trained VGG network, and a detailed inspection of the effect of Octave Convolutions and its low-frequency ratio ($\alpha$).}
\label{tab:ablations}
\begin{tabular}{clccccc}
\hline
\multicolumn{2}{c}{Methods}                     & SSIM ($\uparrow$) & PSNR ($\uparrow$) & MAE ($\downarrow$)& MSE ($\downarrow$) & Time ($\downarrow$)            \\ \hline \hline
\multicolumn{2}{c}{Macro2Micro (w/o brainPD)}    & \textbf{0.8663} & \textbf{26.2560} & 0.1440          & 0.1288          & 0.0110          \\
\multicolumn{2}{c}{Macro2Micro (w/o perct.loss)} & 0.8565          & 25.7207          & 0.1407          & 0.1251          & 0.0110          \\
\multicolumn{2}{c}{Macro2Micro}                 & 0.8631          & 26.0478          & \textbf{0.1374} & \textbf{0.1221} & 0.0110          \\ \hline
\multicolumn{2}{c}{No Octave}                   & 0.8598          & 25.8797          & \textbf{0.1280}          & \textbf{0.1122}          & \textbf{0.0056}          \\
\multicolumn{2}{c}{Octave $\alpha=0.25$}                 & \textbf{0.8640} & \textbf{26.1565} & 0.1417 & 0.1264 & 0.0126 \\
\multicolumn{2}{c}{Octave $\alpha=0.50$ (Default)}       & 0.8631 & 26.0478 & 0.1374 & 0.1221 & 0.0110 \\
\multicolumn{2}{c}{Octave $\alpha=0.75$}                 & 0.8597          & 25.9887          & 0.1414          & 0.1258          & 0.0129          \\ \hline
\end{tabular}
\end{table}

\textbf{Effectiveness of Octave Convolution.} \label{sec:result_octave}
We tested how Octave Convolution influences our suggested model. When Octave Convolution was utilized, both the SSIM and the PSNR improved, as demonstrated in Table~\ref{tab:ablations}. The $\alpha$ value in Octave Convolution represents the percentage of low-frequency features relative to total features and verifies the difference. In Figure~\ref{fig:Octave}, at an $\alpha$ of 0.25, it is clear that the low-frequency image has a low image contrast, while the high-frequency image has a high contrast. Low-frequency images reveal more information with an $\alpha$ of 0.75 than at values of 0.5 and 0.25. On the other hand, if $\alpha$ is set to 0.5, it's clear that the image-based separation of high and low frequencies seems balanced. The best results in terms of MAE, MSE, and inference time were achieved with an $\alpha$ value of 0.5, while an $\alpha$ value of 0.25 yielded the higher SSIM and PSNR but achieved similar image quality with an $\alpha$ value of 0.5 (Table~\ref{tab:ablations}).

\begin{figure}[t!]
    \centering
    \includegraphics[width=\textwidth]{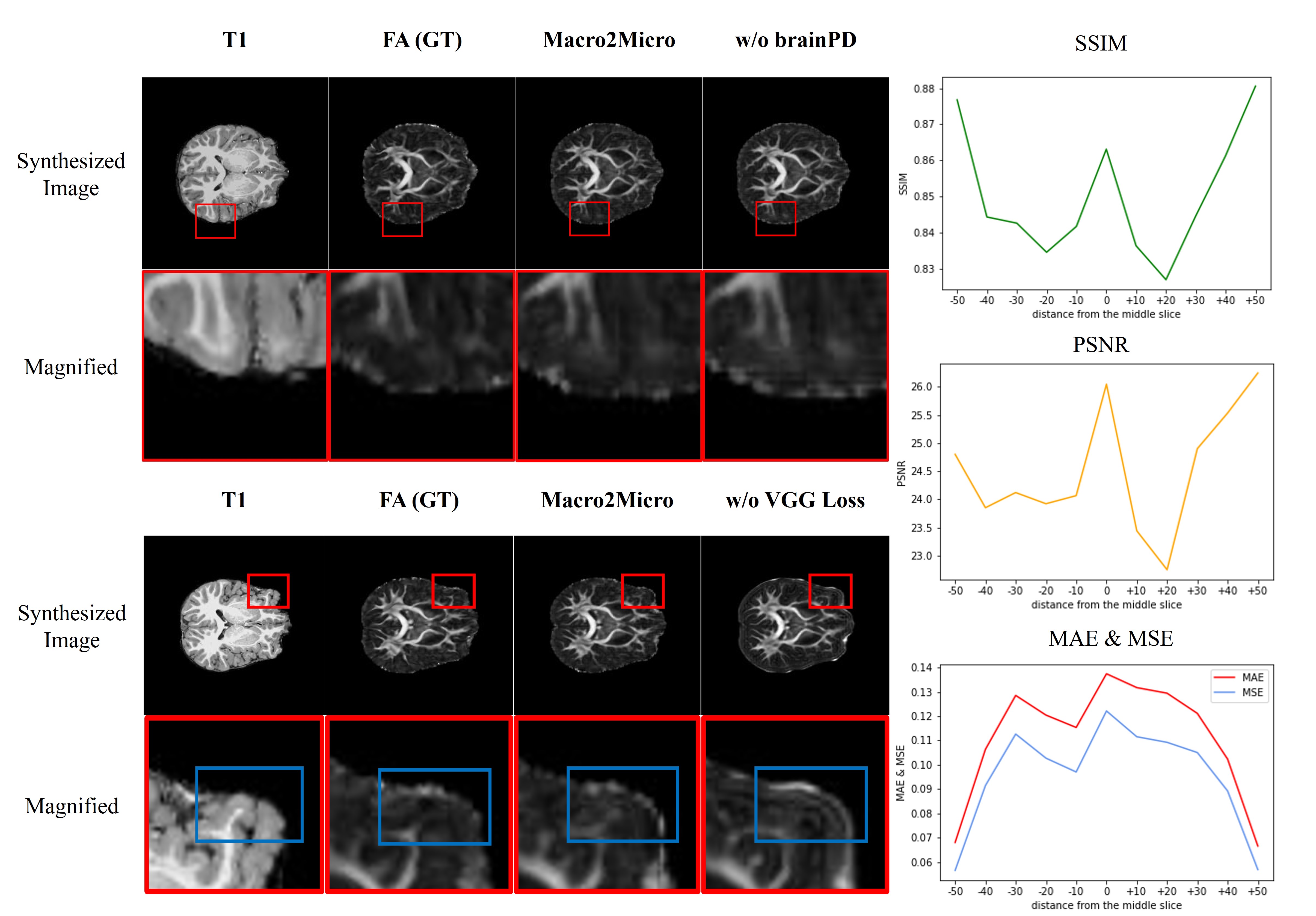}
    \caption{\textbf{(Left)} Generated FA images with and without using a brain-focused patch discriminator and perceptual loss using a pre-trained VGG network. \textbf{(Right)} Inference performance of our model along the distance from the center of the brain.}
    \label{fig:brainPD_percept_loss_distance}
\end{figure}

\textbf{Ablation Studies.} \label{sec:ablations}
Fig.~\ref{fig:brainPD_percept_loss_distance} and Table.~\ref{tab:ablations} show the effectiveness of design choices employed in our model. For brain-focused patch discriminator (brainPD), a model without brainPD generates artifacts like white dots along the boundaries of the brain, and the checkerboard patterns are generated. As the brainPD focuses on the effective regions of the brain, the model with brainPD synthesizes the target modality in higher image quality with more delicate details and generates fewer artifacts. Similarly, a model trained without perceptual loss using a pre-trained VGG network shows lower performance in all evaluation metrics, generating severe artifacts that look like a brain skull at the boundary of the brain (blue rectangles). 

Although our method was trained using only the central slice of the brain, it worked well not only in the center but also in its periphery. Figure~\ref{fig:brainPD_percept_loss_distance} depicts how each evaluation metric shifts from the center to its periphery. Our model performs strongly in both the peripheral and central slices, showcasing its exceptional generability. It is worth noting that the score is higher towards the extreme end of the brain than in the center. We believe this is because the brain size in the image itself in the extreme end of the brain is smaller compared to the central slice. Therefore, the metric includes more backgrounds, resulting in improved performance.

\begin{figure}[t]
    \centering
    \includegraphics[width=\textwidth]{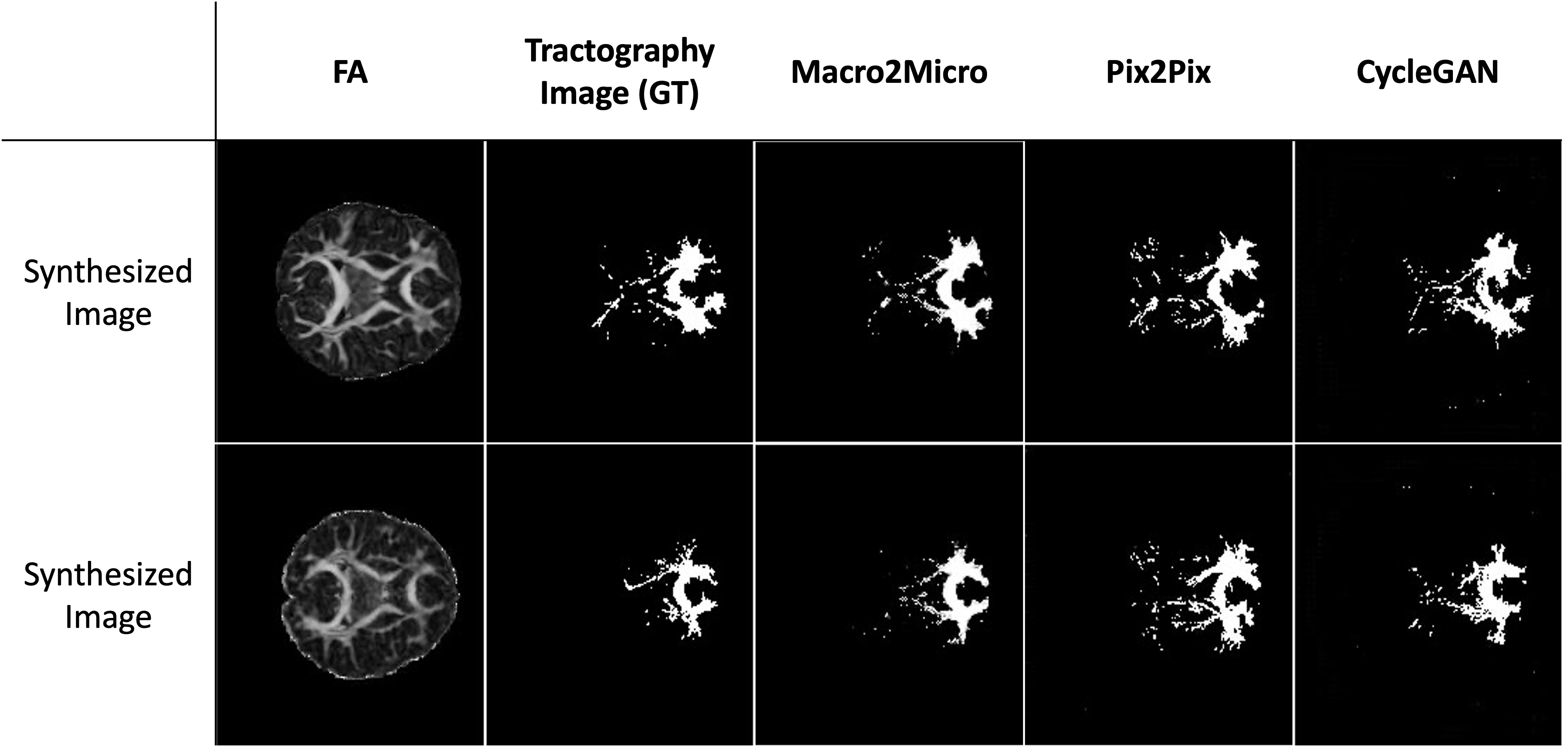}
    \caption{Qualitative comparison of generated 2D Tractography from FA images.}
    \label{fig:phase2}
\end{figure}

\textbf{FA Image Translation to Tractography.} Our model's results were most comparable to the ground truth tractography. In Figure~\ref{fig:phase2}, the overall position of the fiber passing through the vmPFC region and the dense region where many streamlines pass were well predicted. Our results exhibit less noise and achieve more accurate generation of target regions compared to Pix2Pix and CycleGAN, which tend to yield noisier results around the target tractography. Additionally, our results demonstrate state-of-the-art performance in terms of SSIM, PSNR, MAE, and MSE scores, as shown in Table~\ref{tab:phase2}.

\begin{table}[t!]
  \small \centering
  \caption{Quantitative Comparison of generated tractography images with baseline models.}
  \label{tab:phase2}
  \renewcommand{\arraystretch}{1.1}
  \resizebox{\columnwidth}{!}{
  \begin{tabular}{ccccc}
    \toprule
    \hline
    Methods & SSIM ($\uparrow$) & PSNR ($\uparrow$) & MAE ($\downarrow$)& MSE($\downarrow$) \\ \hline
    \midrule 
    Pix2Pix & 0.8206 & 15.0687 & 0.0654 & 0.0620 \\
    CycleGAN & 0.8048 & 15.7659 & 0.1270 & 0.1236 \\ \hline
    Macro2Micro & \textbf{0.8551} & \textbf{17.0572} & \textbf{0.0521}& 
    \textbf{0.0474} \\
    \hline
    \toprule 
    \end{tabular}}
\end{table}

\section{DISCUSSION}
The current discoveries will hold significance for both clinical applications and research. For instance, since the inference time for our model takes less than 0.011 seconds, Macro2Micro can substantially decrease the time and expenses needed to acquire many MRIs clinically. This can be efficiently utilized for newborns and individuals with illnesses incapable of undergoing time-consuming MRI procedures. For example, T1-weighted MRI typically requires five minutes to complete, DTI takes approximately 720 minutes per patient, and tractography takes over 12 hours per subject\cite{hagler2019image}. In contrast, Macro2Micro can offer a cross-modal MRI synthesis in less than a minute at the latest. From a research perspective, researchers can generate a dataset that is essential for diagnostic or research purposes in a multi-modal fashion. Overall, Macro2Micro is crucial in enhancing the quality and efficiency of cross-MRI synthesis tools in both research and clinical settings.

In this study, we propose a new image-to-image translation framework based on the generative adversarial network, Macro2Micro, for predicting microstructure of the brain from the macrostructure of the brain. The human brain is a complex system consisting of several components, spanning from individual cells to interconnected circuits and networks. It demonstrates the characteristics of a complex system in which emergent qualities become evident when one transitions from microscale to macroscale. Thus, determining the brain's structural connectivity from macrostructure to microstructure is critical to our understanding of its functional dynamics in cognitive tasks and how each component affects the developmental processes of childhood and adolescence. From the clinical perspective, obtaining multi-scale data non-invasively has been difficult due to lack of data and costly and time-consuming acquisition process of MRIs\cite{jones2004effect,jones2013white,lueken2011don,murphy1997adult,rzedzian1983real,tsao2010ultrafast,aksoy2008single}. 

Moreover, unlike natural images that were widely explored in conventional image-to-image translation tasks, brain images are grayscale, which have one channel to represent their value. Thus, one should focus more on structural information, contrast, and edges rather than color information when it comes to adopt image-to-image translation techniques with the brain modalities. Considering this fact and different spatial characteristics of the macro and microstructure of the brain, we utilized Octave Convolutions to encode the input modality and decode the target modality. Octave Convolutions encode the structural features of the brain's various hierarchical layers based on its spatial frequencies without transforming the brain modality into a spectral domain. In addition, an active information exchange occurring between two frequency processing branches in both encoding and decoding process promotes the model to capture structural connectivity between different MRI modalities.

In most cases, the brain images include a significant portion of the redundant background. This impedes the entire learning of the model in that this guides the model to concentrate on relatively unrelated regions rather than the actual brain. Furthermore, these redundant "zero" values in the backgrounds are vulnerable to be detected by the discriminator and thus easily result in outputs with inferior quality, leading to the mode collapse\cite{park2020swapping}. We tackled this by enlarging the brain and cropping it into patches for a focused analysis with our brain-focused patch discriminator. The brain-focused patch discriminator extracts patches from images and determines whether each patch belongs to the actual brain data or not. This approach helps to alleviate the challenges posed by the background dominance and allows the model to better capture and generate high-quality brain images with improved details, accuracy in brain region representation, and reduced artifacts.

We further utilized the prior knowledge from the pre-trained convolutional neural networks to guide our model to additionally learn brain-agnostic representations. This methods resulted in superior output with fine details and more accurate brain boundaries in the generated images, effectively eliminating undesired artifacts such as skull-shaped patterns near brain boudaries. This suggests that, on some level, generic image data and brain images may share image features. This finding indicates the potential for extending pre-trained models from general data to the brain domain, opening up possibilities for further exploration.

We validated the capability of the proposed model to encode and generate Fractional Anisotropy (FA) images containing micro-structure information from T1-weighted MRI (T1), which measures the macro-structure of the brain. Various metrics that evaluate the quality of generated images are utilized for quantitative assessment. The results reveal the superior performance of Macro2Micro over widely used image-to-image translation models such as CycleGAN\cite{zhu2017unpaired} and Pix2Pix\cite{isola2017image} both qualitatively and quantitatively. 

Our model preserves the structural information from T1 to produce an FA image, resulting in an output that not only shows a high level of agreement with the ground truth modality but often surpasses it with the finest structural details. Macro2Micro precisely builds detailed white matter structures and fills in gaps in areas where the ground truth FA shows unconnected or distorted regions. This suggests that our model is capable of understanding the relationship between macro-scale and micro-scale structures, enabling it to gather extra information from T1 that is not included in the reference FA image and integrate them. As a result, our model could generate small white matter structures with higher accuracy. It illustrates the usefulness of macro-scale knowledge in building micro-structures, in addition to micro-scale information. Furthermore, feature reduction method such as principal component analysis (PCA) is applied to evaluate the similarity in low-dimensional image representations between the FA images generated from T1 images and real FA images, enhancing the explainability of generated results.

While generated FA images from Macro2Micro visually resembled the ground truth FA images, their clinical value hinges on retaining critical individual differences and there exists a risk of compromising biological information from the actual T1 images during the translation process\cite{cohen2018distribution}. In such cases, the clinical application would be significantly limited. Thus, we executed downstream classification tasks to evaluate whether generated FA images lose individual biological characteristics while being transferred from input T1 images. We conducted a task of predicting the sex, intelligence, and Attention-Deficit/Hyperactivity Disorder (ADHD) diagnosis of children using generated and ground truth FA images both in two dimensional(2D) space and three dimensional(3D) space. Macro2Micro exhibits higher performance in predicting sex, intelligence and ADHD compared to actual FA image in 2D space, and achieved comparable qualities in 3D space. Given the fact that our model was trained using only the central slice of the brain to make an entire 3D brain, our model shows superior performance on generating the target modality while retaining fundamental individual biological features. The heightened phenotype prediction performance of generated FA images might be attributed to their resilience against scanner effects. Training the model on multiple subjects can enhance its capacity to establish a consistent mapping between macrostructure and microstructure.

This robustness to versatile factors contributes to the emergence of more prominent biological characteristics in the generated images. This suggests that our model operated in a manner that preserved or even amplified biological information during the transformation from T1 to FA, and using Macro2Micro is highly practical in that this can provide further biological information (e.g., sex, the presense of ADHD, intelligence) beyond generating the target images themselves. Evaluation of generated images using such approaches has not been frequently conducted in previous image generative modeling studies. If future studies extend their research to variables beyond sex, the presence of ADHD, and intelligence, the utilization of generated images can provide researchers with various valuable benefits. 

the previous studies \cite{Yang2020CMI2I} have been limited to the task of transforming between two brain modalities in two dimensions, our study demonstrates the potential of predicting the target whole brain modalities in three dimensions with the model trained only using the central slice of the input brain modality. Our method provides the effective mapping of nonlinear relationships between brain structures from macrostructure to microstructure using the generative deep learning model, offering a new perspective in understanding the multi-scale structure connectivity of the brain.

There are some limitations of this study.  In our model, only a single slice positioned in the middle along the z-axis was used for training. Consequently, the performance at the ends of the z-axis was relatively lower than that at the central regions. This is because the images at the ends of the axis contain too large portion of redundant backgrounds that even our brain-focused discriminator cannot manage. This can be alleviated if additional learning objectives are applied or through retraining the model with the images at the ends of the axis. In future research, incorporating all slices of the brain and training the model could lead to better outputs even at the peripheral regions.

On the other hands, although we showed the applications of our model to predict tractography from given DTI and demonstrated superior performance against other baseline models, there were still noticeable differences compared to actual tractography images both in 2D and 3D. Nevertheless, since we have proven in this study that tractography can be created from macrostructures of the brain through our model, we expect that we will be able to create more efficient, accurate and sophisticated images through further research.

In summary, this research successfully generated microscopic structures from macrostructures of the human brain, demonstrating the interrelation between brain structures at different scales. Furthermore, it effectively learned subject-independent structures by accommodating individual variations. The current findings will be important both for clinical applications and research. In clinical practice, Macro2Micro has the potential to significantly reduce the time and cost required to obtain Diffusion Tensor Imaging (DTI). Diffusion MRI is the most sensitive technique for diagnosing acute cerebral infarction and can help determine treatment for hyperacute cerebral infarction by showing lesions within a few minutes of occurrence. However, in time-critical clinical trials, it takes a lot of time to capture images of both Diffusion MRI and structural MRI(sMRI). If we take a single sMRI and use it to create a corresponding DTI with our model, the time for diagnosis will be significantly reduced. This can be effectively used for infants and patients with anxiety disorders who are unable to undergo time-consuming MRI procedures. In research practice, by generating data from existing patients with sMRI, one can create dataset needed for diagnosis or research in a multi-modal manner.

\bibliographystyle{splncs04}
\bibliography{reference}
\end{document}